\documentclass[aps,showpacs,superscriptaddress,showkeys,nofootinbib,floatfix]{revtex4}
\usepackage{graphicx}

\def\slashchar#1{\setbox0=\hbox{$#1$}
   \dimen0=\wd0 \setbox1=\hbox{/} \dimen1=\wd1
   \ifdim\dimen0>\dimen1 \rlap{\hbox to \dimen0{\hfil/\hfil}} #1
   \else  \rlap{\hbox to \dimen1{\hfil$#1$\hfil}} / \fi}

\newcommand{\comp}{{\rm C}\hspace{-1ex}\rule{0.1mm}{1.5ex}\hspace{1ex}}

\begin{document}

\title{Properties of the $\rho$ and $\sigma$ Mesons from Unitary
  Chiral Dynamics.}
\author{J.~Nieves}\email{jmnieves@ific.uv.es}
\affiliation{Instituto de F{\'\i}sica Corpuscular (centro mixto CSIC-UV)\\
Institutos de Investigaci\'on de Paterna, Aptdo. 22085, 46071, Valencia, Spain
} 
\author{E. Ruiz
  Arriola}\email{earriola@ugr.es}\affiliation{Departamento de
  F\'{\i}sica At\'omica, Molecular y Nuclear, Universidad de Granada,
  E-18071 Granada, Spain.}

\date{\today}

\begin{abstract} 
\rule{0ex}{3ex} The chiral limit of the $\rho$ and $\sigma$ masses and
widths is discussed. We work within the inverse amplitude method to
one loop in SU(2) ChPT and analyze the consequences that all chiral
logarithms cancel out in the $\rho-$channel, while they do not cancel
for the $\sigma$ case, and how they strongly influence the properties
of this latter resonance. Our results confirm and explain the
different behavior of the $\sigma$ and $\rho$ poles for $N_C$ not far from
3, but we extend the analysis to very large $N_C$, where the behavior
of these two resonances is re-analyzed. We note that the rather natural
requirement of consistency between resonance saturation and
unitarization imposes useful constraints. By looking only at the
$\rho-$channel, and within the single resonance approximation, we find
that the masses of the first vector and scalar meson nonets, invoked
in the single resonance approximation, turn out to be degenerated in
the large $N_C$ limit. On the contrary we show that, for sufficiently
large $N_C$, the scalar meson evolution lies beyond the applicability
reach of the one-loop inverse amplitude method and if the scalar
channel is also incorporated in the analysis,  it may lead, in some
cases, to phenomenologically inconsistent results.
\end{abstract}

\maketitle



\section{Introduction}

The large $N_C$-limit of QCD~\cite{'tHooft:1973jz,Witten:1979kh} makes
quark-hadron duality manifest at the expense of introducing an
infinite number of weakly interacting stable mesons and
glueballs. While the corresponding counting rules are deduced at the
quark-gluon level, internal consistency requires them to be valid also
at the hadron level. Moreover, large $N_C$ studies might clarify
several features of the nuclear force, in particular, the role played
by the ubiquitous scalar meson. This is an essential ingredient which
contributes to the mid range nuclear attraction, which, with a mass of $
\sim 500$ MeV, was originally proposed in the
fifties~\cite{PhysRev.98.783} to provide saturation and binding in
nuclei.  During many years, there has been some arbitrariness on the
``effective'' scalar meson mass and coupling constant to the nucleon,
partly stimulated by lack of other sources of information. For
instance, in the very successful nucleon--nucleon charge dependent 
Bonn potential~\cite{Machleidt:2000ge} any partial wave $^{2S+1}
L_J$-channel is fitted with noticeably different scalar meson masses
and couplings. The situation has steadily changed during the last
decade, and the scalar meson has been finally
resurrected~\cite{Tornqvist:1995ay}, culminating with the inclusion of
the $0^{++}$ resonance in the PDG~\cite{Yao:2006px} as the $f_0 (600)$
resonance, also denoted as the $\sigma$.  The $\sigma$-resonance is
traditionally seen in $\pi\pi$ scattering, with a wide spread of
values being displayed ranging from $400-1200 $ MeV for the mass and a
$600-1200$ MeV for the width~\cite{vanBeveren:2002mc}. These
uncertainties have recently been sharpened by a benchmark
prediction based on Roy equations and chiral
symmetry~\cite{Caprini:2005zr} yielding the unprecedented accurate
values
\begin{equation}
\sqrt{s_\sigma}= 441^{+16}_{-8} - {\rm i}\ 272^{+9}_{-12}~ {\rm
  MeV}.\label{eq:roy} 
\end{equation}
Forward dispersion relations and Roy equations on the real axis have also
been used by the Madrid group in Ref.~ \cite{Kaminski:2006qe} yielding a
slightly heavier and narrower $\sigma$-resonance determination.

While the existence of this broad low lying state is by now out of
question, the debate on the nature of the $\sigma$-meson is not
completely over. Structures of the type tetraquark or glueball,
etc. have been proposed (see e.g. Ref.~\cite{Klempt:2007cp} for a
recent review and references therein).

Lattice determinations of the lightest scalar mesons are challenging
(for reviews see~\cite{Prelovsek:2008qu,Liu:2008ee}). It has been
found that the mass of the lightest $0^{++}$ meson is suppressed
relative to the mass of the $0^{++}$ glueball in quenched QCD at an
equivalent lattice spacing~\cite{Hart:2006ps}.  In the quenched
approximation it has been claimed~\cite{Mathur:2006bs}, that $m_\sigma
= 550~ {\rm MeV}$ for pion masses as low as $m_\pi=180~ {\rm MeV}$. On
the other hand, a recent analysis of $np$ scattering in the $^1S_0$
channel yields $m_\sigma= 510(10)~ {\rm
MeV}$~\cite{CalleCordon:2008eu,CalleCordon:2008cz}, when uncorrelated
$2\pi$ exchange is disregarded (see however
Refs.~\cite{Oset:2000gn,Jido:2001am}).

Chiral Perturbation Theory (ChPT) for $\pi\pi$ scattering has been
introduced in Refs.~\cite{Lehmann:1972kv,Gasser:1983yg}. We aim here
to re-analyze the nature of the ChPT scalar resonance, stressing its
differences and similarities with the $\rho-$meson. Most of the
results will be obtained in the limit of massless pions, which will
allow us to work with almost analytical equations, hence simplifying
and enlightening the discussion. We will use a suitable generalization
of the effective range expansion~\cite{Lehmann:1972kv} such as the
Inverse Amplitude
Method~\cite{Truong:1988zp,Dobado:1989qm,Dobado:1996ps,Hannah:1997ux}
to unitarize the one loop $\pi\pi$ ChPT amplitudes. The method has
been successfully used in the past not only for $\pi\pi$ (coupled
channel meson--meson in general)~\cite{Dobado:1992ha,Oller:1997ng,
GomezNicola:2001as, Nieves:2001de,GomezNicola:2007qj}, but also for
$\pi N-$scattering~\cite{GomezNicola:2000wk,Nicola:2003zi}. Poles of
the unitarized amplitudes in unphysical sheets provide masses and
widths of the resonances, and the change of the position of these
poles determines their properties for growing $N_C$. To carry out such
a program requires extending the chiral amplitudes for an arbitrary
number of colours. This is a delicate point because in any case one
must insure that {\it all} possible leading $N_C$ dependences should
be contained in this unphysical $N_C > 3 $ extrapolation. In our study
all QCD $N_C$ dependence appears through the Low Energy Constants
(LEC's), which leading $N_C$ behaviour could be obtained within the
resonance saturation approach~\cite{Ecker:1988te,Ecker:1989yg}. We
will further simplify the problem and we will work within the Single
Resonance Approximation (SRA) scheme, where each infinite resonance
sum is just approximated with the contribution from the lowest-lying
meson nonet with the given quantum numbers. We will see how a rather
natural requirement of consistency between resonance saturation and
unitarization imposes useful constraints in the extreme $N_C \to
\infty$ limit.

The previous works of Refs.~\cite{Pelaez:2003dy, Pelaez:2006nj} use a
similar framework, however with a different $N_C$ scaling strategy;
the $N_C=3$ fitted results are simply re-scaled to the unphysical
$N_C$ values. One of the major, but crucial, differences of the present
work with these other two is the use of the SRA to estimate the
leading $N_C$ behaviour of the LEC's. Though in the vicinity of
$N_C=3$ we find similar results, we will show that the requirement of
consistency between resonance saturation and unitarization, for very
large values of $N_C$, imposes useful constraints.

\section{One Loop $\pi\pi$ ChPT Amplitudes in the Chiral Limit}

After projecting out in isospin and angular momentum ($IJ$), the
scalar--isoscalar and vector-isovector $\pi\pi$ amplitudes to one loop
accuracy and in the chiral limit
read~\cite{Lehmann:1972kv,Gasser:1983yg} (in the centre of mass
frame):
\begin{eqnarray}
T^{00} (s)&=&   
\underbrace{-\frac{s}{f^2}}_{T^{00}_2} ~ \underbrace{- \frac{s^2}{576\pi^2f^4} \left \{28 {\bar l_2} + 22 {\bar l_1} + 51 -14 \ln (s/m^2) -36 \ln(-s/m^2)
\right\}}_{T^{00}_4} \label{eq:t00}\\\
T^{11} (s)&=& 
 \underbrace{-\frac{s}{6f^2}}_{T^{11}_2} ~ \underbrace{-\frac{s^2}{576\pi^2f^4} \left \{2{\bar l_2} - 2 {\bar l_1} + \frac23 + \ln (s/m^2)- \ln(-s/m^2)
\right\}}_{T^{11}_4} \label{eq:t11}
\end{eqnarray}
where $s$ is the square of the total energy of the two pions, $f$ is
the pion decay constant in the chiral limit ($\sim$ 88 MeV), $m $ is
the pion mass (this apparent dependence on $m$ is fictitious, 
as we will see below) and $T_2^{IJ}$ and $T_4^{IJ}$ are the tree
level and one loop amplitudes, and the normalization of the total
amplitude is fixed by its relation with the elastic phase shifts
\begin{equation}
T^{IJ} = -8\pi \sqrt{s} \left ( \frac{e^{2{\rm i}\delta^{IJ}}-1}{2{\rm i} p} \right)
\end{equation}
with $p$ the centre of mass pion  momentum, $\sqrt{s}/2$ for massless pions. 
The logarithm in the above equations is defined as ($z\in \comp$)
\begin{equation}
\ln z = \ln |z| + {\rm i~Arg}(z), \quad {\rm Arg}(z)
\in [-\pi, \pi[
\end{equation}
In Eqs.~(\ref{eq:t00}) and (\ref{eq:t11}) the last logarithm
($\ln(-s/m^2)$, with $m$ the pion mass) produces the unitarity right
hand cut and it accounts for perturbative two particle elastic
unitarity
\begin{equation}
{\rm Im}T_4^{IJ}(s+{\rm i}0^+) = 
-\frac{|T_2^{IJ}(s)|^2}{16\pi} + {\cal O}(1/f^4), \quad s>0
\end{equation}
while the first logarithm in these two equations provides the left 
hand cut required by crossing symmetry, and it leads to complex
amplitudes for $s<0$. Finally, ${\bar l}_{1,2}$ are scale independent
LEC's. Up to a numerical factor, the quantity
${\bar l}_i$ is the value of the renormalized coupling constant
$l_i^r(\mu)$ at the scale $\mu=m$,
\begin{equation}
{\bar l}_1 = 96 \pi^2 l_1^r(\mu) - \ln(m^2/\mu^2), \quad {\bar l}_2 =
48 \pi^2 l_2^r(\mu) - \ln(m^2/\mu^2)
\end{equation}
The LEC's ${\bar l}_i$ do not exist in the chiral limit, $m \to 0$,
but contain a chiral logarithm with unit coefficient, i.e., in the
chiral limit ${\bar l}_i$ tend to infinity like $-\ln m$. Note
however, all dependence on the pion mass $m$ cancels out in
Eqs.~(\ref{eq:t00}) and (\ref{eq:t11}), as expected, since $T^{00}$
and $T^{11}$ are well defined in the chiral limit $m\to 0$. The one
loop amplitudes depend on a certain scale $\mu$ through the
renormalized LEC's $l_i^r(\mu)$ and the right (unitarity) and left 
hand cut logs $\ln(-s/\mu^2)$ and $\ln(s/\mu^2)$, respectively.

Already at this level we note here, the remarkable difference between
the $\sigma-$ and $\rho-$meson channels which becomes obvious from the
effective range expansion in the early work of
Lehmann~\cite{Lehmann:1972kv}. For $s>0$, and besides the
imaginary part, left and right hand cut logs cancel out in the
$\rho-$meson channel, while these logs add up in the scalar--isoscalar
sector. Moreover, at a scale of about 770 MeV, the contribution of the
logs is comparable in size to that of the renormalized LEC's $l_i^r$,
being thus the $\sigma-$channel dynamics strongly influenced by
these logarithms stemming from the chiral loops.

Away from the chiral limit, some logs survive in the $\rho-$channel as
well, but their contribution is suppressed by powers of the pion of
mass, i.e. terms of the type $m^2  \ln(\pm s/m^2)$ (see for instance
the appendix B of Ref.~\cite{Nieves:1999bx}, where analytical
expressions for the left hand cut integrals can be found).

Finally, we recall here the relation among the SU(2)$\times$SU(2) and
SU(3)$\times$SU(3) LEC's~\cite{Gasser:1984gg},
\begin{equation}
l_1^r(\mu) = 4 L_1^ r(\mu) + 2L_3 - \frac{\nu_K}{24}, \quad l_2^r(\mu)
= 4 L_2^ r(\mu)  - \frac{\nu_K}{12}
\end{equation}
with $32\pi^ 2\nu_K=\left(\ln(\bar m^2_K/\mu^2)+1\right)$, with $\bar
m_K\sim 468$ MeV the kaon mass in the $m\to 0$ limit, which permits
re-write the amplitudes in terms of $L_{1,2,3}^r(\mu)$.  The scale
dependence of the SU(3)$\times$SU(3) LEC's reads
\begin{equation}
L_i^r(\mu_2) = L_i^r(\mu_1) + \frac{\Gamma_i}{(4\pi)^2} \ln(\mu_1/\mu_2), \quad 
2\Gamma_1=\Gamma_2= \frac{3}{16}, \,\Gamma_3=0. \label{eq:L-ren}
\end{equation}

\section{One Loop IAM $\rho$ and $\sigma$ Poles}

Any unitarization method resums a perturbative expansion of the
scattering amplitude in such way that two body elastic unitarity
\begin{equation}
{\rm Im}\left(\frac{1}{T^{IJ}}\right) = \frac{p}{8\pi\sqrt s}, \quad s> 4m^2
\end{equation}
is implemented exactly. Let us  pay
attention to the one loop IAM, in which the $T^{IJ}-$matrix is approximated
by~\cite{Truong:1988zp,Dobado:1989qm,Dobado:1996ps,Hannah:1997ux}
\begin{eqnarray}
T^{IJ}(s) = \frac{(T_2^{IJ})^2(s)}
{\phantom{\Big(} T_2^{IJ}(s)-T_4^{IJ}(s)} \, ,  \label{eq:iam}
\end{eqnarray}
which perturbatively reproduces ChPT to one loop\footnote{Some
problems associated to the exact position of the Adler's zeros and the
reliability of the IAM predictions for scalar waves have been recently
discussed in Ref.~\cite{GomezNicola:2007qj}.}. Resonances correspond
to poles in the fourth quadrant of the Second Riemann Sheet (SRS),
defined by continuity on the upper lip of the right unitarity cut with
the physical First Riemann Sheet (FRS),
$T_{\rm SRS} (s \pm i 0^+) = T_{\rm FRS} (s\mp i 0^+) $. Thus within this
scheme, we find that mass and width of the resonance ($s_R= m^2_R - i
m_R \Gamma_R$) are determined from the zeros of the denominator
of Eq.~(\ref{eq:iam}) in the SRS,  
\begin{equation}
T_2^{IJ}(s_R)=T_4^{IJ}(s_R)\Big|_{\rm SRS}\, .
\end{equation}
The above condition  leads to  simple equations for the $\rho$ and
$\sigma$ resonances:
\begin{eqnarray}
s_\rho &=& \frac{288 \pi^2 f^2}{\phantom{\Big(} 3{\rm i}\pi +2 +6{\bar l}_2-6{\bar l}_1} = \frac{96 \pi^2 f^2}{\phantom{\Big(} {\rm i}\pi +2/3 -384\pi^2\left(2L_1^r(\mu)-L_2^r(\mu)+L_3\right) } \label{eq:srho} \\
s_\sigma &=& \frac{288 \pi^2 f^2}{\phantom{\Big(} 18{\rm i}\pi+25{\rm i~arctan}\left(\frac{\Gamma_\sigma}{m_\sigma}\right)-25\ln \left|s_\sigma/m^2 \right| +51/2 +14{\bar l}_2+11{\bar l}_1} \nonumber\\
&=&\frac{288 \pi^2 f^2}{\phantom{\Big(} 18{\rm i}\pi+25{\rm i~arctan}\left(\frac{\Gamma_\sigma}{m_\sigma}\right)-25\ln \left|s_\sigma/\mu^2 \right| +51/2 
+ 192\pi^2 \left(22L_1^r(\mu)+14 L^r_2(\mu)+11 L_3 -\frac{25}{48}\nu_K\right)} \label{eq:ssigma}
\end{eqnarray}
where to compute the  amplitude  in the fourth quadrant of the 
SRS at $s=s_R$, we have used:
\begin{eqnarray}
\ln(-s_R)&=&\ln\left(-m^2_R+{\rm i} m_R\Gamma_R\right) = \ln|s_R|+{\rm i}\left[\pi-{\rm arctan}\left(\frac{\Gamma_R}{m_R}\right)\right]
 - 2\pi {\rm i} \label{eq:srs1} \\
\ln(s_R)&=&\ln\left(m^2_R-{\rm i} m_R\Gamma_R\right) = \ln|s_R|-{\rm
  i~arctan}\left(\frac{\Gamma_R}{m_R}\right) \label{eq:srs2}
\end{eqnarray}
We note that in Eqs.~(\ref{eq:srho}) and~(\ref{eq:ssigma}), large
chiral logarithms of the pion mass do not appear, which guaranties
mild pion mass dependences of the $\rho-$ and $\sigma-$masses and
widths~\cite{Hanhart:2008mx}. This is highly desirable to better
control the needed chiral extrapolation of lattice QCD calculations.

From Eq.~(\ref{eq:srho}), we trivially find 
\begin{equation}
   m^2_\rho = 48\pi^2f^2\frac{\bar l_2-\bar l_1 +
   1/3}{\phantom{\Big(}(\bar l_2-\bar l_1 + 1/3)^2+\pi^2/4}, \quad
   \Gamma_\rho = \frac{\pi}{2} \frac{m_\rho}{\phantom{\Big(}\bar
   l_2-\bar l_1 + 1/3}  \label{eq:rho-prop}
\end{equation}
(we express the $\rho-$mass and width in terms of the combination
$\bar l_2-\bar l_1=-192\pi^2\left(2L_1^r(\mu)-L_2^r(\mu)+L_3\right)$,
because in this difference the logarithm of the pion mass cancels
out). The above expressions give reasonable estimates of the pole
position of the $\rho-$resonance.  Using, for instance, at the scale
$\mu=m_\rho\sim 770$ MeV, the set of LEC's determined in
Ref.~\cite{Amoros:2000mc} from an ${\cal O}(p^6)$ study of the
$K_{\ell 4}$ decays,
\begin{equation}
10^3\ L_1^r(m_\rho)=0.52 \pm 0.23, \quad 10^3\ L_2^r(m_\rho)=0.72 \pm 0.24, \quad 10^3\ L_3=-2.70 \pm 0.99 \label{eq:Lr's}
\end{equation}
and taking into account the strong correlations among the $L_i^r$, we
estimate~\cite{Nieves:1999zb} $\bar l_1=0.3 \pm 1.2$ and $\bar
l_2=4.77 \pm 0.45$, with a linear correlation coefficient $r(\bar l_1,
\bar l_2)=-0.69$, that leads\footnote{For narrow resonances, their
mass and width are related to the pole position as $\sqrt{s_R}\approx
m_R-{\rm i}\ \Gamma_R/2$, and it is also usual to use this notation
for broader resonances. The use of $\sqrt{s_R}$ to determine the
resonance mass and width leads to heavier and narrower states. As
a matter of example, with the parameters of Eq.~(\ref{eq:Lr's}), we find
$\sqrt{s_\rho} = \left(840^{+140}_{-100}-{\rm i}\
133^{+89}_{-43}\right)\ {\rm MeV}$.  Variations are much more drastic
for the case of the $\sigma-$resonance, because it is significantly
broader than the $\rho-$meson.}  to $m_\rho=830^{+120}_{-90}$ MeV and
$\Gamma_{\rho}=270^{+190}_{-90}$ MeV. Similar results ($m_\rho \sim
815$ MeV and $\Gamma_\rho \sim 254$ MeV) are obtained by using $\bar
l_1=-0.4\pm 0.6$ and $\bar l_2=4.3\pm 0.1$ obtained from an ${\cal
O}(p^6)$ Roy equation analysis of $\pi\pi$
scattering~\cite{Colangelo:2001df}.

Since the analysis carried out here involves only ${\cal O}(p^4)$
amplitudes, one might think, it would be more appropriate to use
values for the LEC's determined from ${\cal O}(p^4)$ accuracy studies.
If we had used  for the central values of $L^r_{1,2,3}(m_\rho)$, the results
from the ${\cal O}(p^4)$ fit of Ref.~\cite{Amoros:2000mc},
\begin{equation}
10^3\ L_1^r(m_\rho)=0.46, \quad 10^3\ L_2^r(m_\rho)=1.49, \quad 10^3\
L_3=-3.18 \label{eq:Lr2's}
\end{equation}
while keeping the same errors and correlations as in
Eq.~(\ref{eq:Lr's}), we would have found $\bar l_1=-0.9 \pm 1.2$,
$\bar l_2=6.23 \pm 0.45$ with $r(\bar l_1, \bar l_2)=-0.69$. Those
values will lead to $m_\rho=690^{+80}_{-60}$ MeV and
$\Gamma_{\rho}=150^{+60}_{-40}$ MeV, in better agreement with data. Of
course, there will be finite pion mass corrections, though we expect
them to be quite small for the $\rho$ mass, while its width will be
reduced by about 25\% ~\cite{Hanhart:2008mx}. Note that if we fix
$(\bar l_2-\bar l_1 + 1/3)$ to $\left(1.25\times
(\Gamma_\rho/m_\rho)_{\rm exp} \times 2/\pi\right)^{-1} \sim 6.45$, as
deduced from the second relation of Eq.~(\ref{eq:rho-prop}) and the
hypothesis that the $\Gamma_\rho/m_\rho$ ratio increases by about 25\%
for massless pions, the first relation of Eq.~(\ref{eq:rho-prop})
leads to $m_\rho \approx 8.33 f$, in excellent agreement with the
experimental value.

Let us now pay attention  to the case of the $\sigma-$resonance. Solving
numerically Eq.~(\ref{eq:ssigma}), with the set of parameters of
Eqs.~(\ref{eq:Lr's}) and (\ref{eq:Lr2's}), we find
\begin{eqnarray}
\sqrt{s_\sigma} &=& \left(401^{+12}_{-16}-{\rm i}\
277^{+23}_{-26}\right)\ {\rm MeV}, \quad {\rm LEC's~from~} {\cal
O}(p^6)~K_{\ell 4}~[{\rm Eq.~(\ref{eq:Lr's})}]
\label{eq:sig1} \\ 
\sqrt{s_\sigma} &=& \left(410^{+8}_{-13}-{\rm i}\
257^{+25}_{-27}\right)\ {\rm MeV}, \quad {\rm LEC's~from~} {\cal
O}(p^4)~K_{\ell 4}~[{\rm Eq.~(\ref{eq:Lr2's})}]
\label{eq:sig2}
\end{eqnarray}
which are in fair agreement with  the 
benchmark determination based on Roy equation and
chiral symmetry~\cite{Caprini:2005zr} given in 
Eq.~(\ref{eq:roy}). Finite pion mass corrections
produce  a moderate enhancement (around 10\%) of the resonance mass,
while its width is reduced by a similar amount~\cite{Hanhart:2008mx}.

In sharp contrast with the case of the $\rho$ meson, the properties of the
$\sigma-$meson ($f_0(600)$) turn out to be strongly influenced by the
chiral logarithm $\ln \left|s_\sigma/\mu^2 \right|$. Actually, there
exist large cancellations in the combination of LEC's $192\pi^2
\left(22L_1^r(\mu)+14 L^r_2(\mu)+11 L_3 -\frac{25}{48}\nu_K\right)$
appearing in Eq.~(\ref{eq:ssigma}), and this contribution plays a role
much less important than in the case of the $\rho-$meson, for which
the LEC's and the discontinuity through the unitarity cut determine
mostly its properties. Indeed, if for the $\sigma-$resonance, the
LEC's contribution is neglected, one will find $\sqrt{s_\sigma}= 417 -
{\rm i}\ 236~ {\rm MeV}$.  The comparison of this latter result with
those displayed in Eqs.~(\ref{eq:sig1}) and (\ref{eq:sig2}) shows that
the bulk of the dynamics of the $\sigma-$resonance is not
determined\footnote{On the contrary, for the case of the $\rho-$meson
  neglecting the LEC's contribution leads to unrealistic results:
$s_\rho=288\pi^2f^2/(2+3{\rm i} \pi))$, which gives
$m_\rho=689$ MeV and $\Gamma_\rho= 3245$ MeV, or equivalently
$\sqrt{s_\rho}= 1175 - {\rm i}\ 951~ {\rm MeV}$. } by the LEC's, but
rather by unitarity ($18{\rm i}\pi$), the constant term $51/2$ and the
chiral logarithm $-25\ln\left(s_\sigma/\mu^2\right)$.  This chiral log, due
to both the left and right hand cut contributions, favours smaller
(larger) values of the $\Gamma_\sigma/m_\sigma$ ratio for 
$|s_\sigma|^\frac12$ smaller (larger) than the renormalization scale
$\mu$.  When this contribution is neglected, we find with the set of
parameters of Eq.~(\ref{eq:Lr's}) [Eq.~(\ref{eq:Lr2's})], that the
$\Gamma_\sigma/m_\sigma$ ratio comes out to be around a factor 2.2
[1.6] greater than if the chiral log was considered. Actually, the
pole exists as long as the real part of the denominator in
Eq.~(\ref{eq:ssigma}) remains positive, and it imposes a constraint to
$ \left|s_\sigma/\mu^2\right|$
\begin{eqnarray}
\frac{\Gamma_\sigma}{m_\sigma} &=& 
\frac{18\pi+25\,{\rm
    arctan}\left(\frac{\Gamma_\sigma}{m_\sigma}\right)}{192\pi^2 \left(22L_1^r(\mu)+14 L^r_2(\mu)+11 L_3
-\frac{25}{48}\nu_K\right) +\frac{51}{2} 
-25\ln\left|s_\sigma/\mu^2\right|} > 0 \label{eq:ratio}\\
&\Rightarrow&  25\ln\left|s_\sigma/\mu^2\right| < 192\pi^2 \left(22L_1^r(\mu)+14 L^r_2(\mu)+11 L_3
-\frac{25}{48}\nu_K\right) +\frac{51}{2} 
\end{eqnarray}

After this discussion, it is easy to understand the nomenclature
commonly used in the literature of {\it dynamically generated}
referred to the $\sigma-$resonance. Actually, it is possible
to describe the scalar channel with the leading order plus a cutoff
(or another regularization parameter) playing the role of some
combination of higher order parameters, while for the case of the
$\rho-$resonance the ${\cal O}(p^4)$ LEC's are
needed~\cite{Oller:1997ng,Oller:1997ti, Nieves:1998hp, Nieves:1999bx}.

\section{Large $N_C$ Limit of the One Loop IAM $\rho$ and $\sigma$ Poles}

The large $N_C$ extension of the $\sigma$ and $\rho$ pole positions
obtained in the previous section  is straightforward. We
should just consider that the pion weak decay constant scales as
${\cal O} (\sqrt{N_C}) $, while the LEC's $L_{1,2,3} $ behave as ${\cal
O}(N_C)$, with $L_2- 2 L_1 = {\cal
O}(N_C^0)$~\cite{'tHooft:1973jz,Witten:1979kh}.  The chiral logs, as
well as the renormalization scale dependence of the LEC's
[Eq.~(\ref{eq:L-ren})] are subleading in $1/N_C$. Thus, we trivially
find for $N_C \gg 3$ and massless pions
\begin{eqnarray}
  \widehat m_\rho^2 &=& -\frac{\widehat f^2}{4\widehat L_3}, \qquad\quad\quad\quad
      \widehat\Gamma_\rho  =
\frac{\widehat m_{\rho}^3}{96\pi \widehat f^2} \label{eq:pole-rho}\\ 
  \widehat m_\sigma^2 &=& \frac{3\widehat f^2 }{ 50 \widehat L_2 +22
      \widehat L_3}, 
\qquad \widehat \Gamma_\sigma = \frac{\widehat m_\sigma^3}{16\pi
      \widehat f^2} \label{eq:pole-sigma}
\end{eqnarray}
where the hat over a symbol ($\widehat O$) implies its value in the
$N_C\gg 3$ limit.  Although we expect that the $N_C$ behavior close to
the physical value $N_C=3$ of the $\sigma$ is non $q\bar q$ due to the
chiral logs, for a sufficiently large $N_C$, the above equations,
provided that $-\widehat L_3$ and $(25 \widehat L_2 + 11 \widehat
L_3)$ are positive quantities, show that for both resonances, the mass
scales as ${\cal O}(N_C^0)$, while the width decreases as $1/N_C$.
Thus, in this $N_C$ regime both resonances would follow a $q \bar q $
pattern in the nomenclature of Refs.~\cite{Pelaez:2003dy,
  Pelaez:2006nj}. Nevertheless, the very large $N_C$ pole in
(\ref{eq:pole-sigma}) could be located at a rather different position
from that of $N_C=3$, as we will discuss below. This large $N_C$ pole
might then be interpreted as a sub-dominant $q \bar q $ component of
the $\sigma$ resonance. From a sufficiently large value of $N_C$ on,
such component may become dominant, and beyond that $N_C$ the
associated pole would behave as a $q \bar q $ state, although the
original state only had a small admixture of $q \bar
q$~\cite{Pelaez:2005fd}. Something similar was found at two loops in
\cite{Pelaez:2006nj}, but we are showing here that it is possible also
at one loop. On the other hand, if either $\widehat L_3$ or $(25
\widehat L_2 + 11 \widehat L_3)$ were exactly zero, the pole position
($s_R$) of the associated resonance would grow with $N_C$ as $\widehat
f^2$, for sufficiently large $N_C$.

In any case, we face here the fundamental problem of determining the
values of $f, L_2$ and $L_3$ for unphysical $N_C\ne 3$ values. The separation
between the large $N_C$ leading and subleading parts of the measured
$L_i$ is not possible. In general, one has the scale independent
combination
\begin{eqnarray}
  L_i^r (m_\rho)=L_i^r (\mu) 
+ \frac{\Gamma_i}{(4\pi)^2} \ln (\mu/m_\rho) =  A_i N_C
  + B_i \, \label{eq:eles-nc}
\end{eqnarray} 
where $A_i$ and $B_i $ are $N_C$ independent. Note that only the
$N_C=3$ combination is experimentally accessible. However a meaningful
extension of the chiral amplitudes to an arbitrary number of colors
requires some knowledge of the different $A_i$ and $B_i$ coefficients,
and of course of those appearing in a similar decomposition of the
pion decay constant. This is of particular importance in the scalar
channel due to the large cancellation\footnote{For instance, the
${\cal O}(p^4)$ values of Eq.~(\ref{eq:Lr2's}) leads to $|(25
L_2^r(m_\rho) + 11 L_3)/(11L_3)| = 0.06^{+0.29}_{-0.03}$.}  existing
in the combination $(25 L_2^r(m_\rho) + 11 L_3)$.

In Ref.~\cite{Pelaez:2003dy}, the prescription of scaling $f\to
f\sqrt{N_C/3}$, and $L_i^r(m_\rho) \to L_i^r(m_\rho) (N_C/3)$ for
$i=2,3$ is adopted. There, $2 L_1^r(m_\rho)-L_2^r(m_\rho)$ is kept
constant and an uncertainty on the scale $\mu=0.5-1$ GeV is also taken
into account.  However, as $N_C$ starts significantly deviating from
the physical value $N_C=3$, such prescription might not be accurate
enough. For example, let us consider the case of the
$\sigma-$resonance. As mentioned above, there exists a large
cancellation in the LEC's contribution to the scalar channel, and it
plays a minor role for $N_C=3$. As a result, not only the absolute
value, but also the sign of $(25 L_2^r(m_\rho) + 11 L_3)$ is subject
to sizable uncertainties. Depending on the set of values used, one
finds different signs for this combination of LEC's, which induces
totally different behaviours when the number of colors grows. Thus,
with the ${\cal O}(p^6)$ values of Eq.~(\ref{eq:Lr's}), this
combination is negative, while it turns out to be positive when the
${\cal O}(p^4)$ set of Eq.~(\ref{eq:Lr2's}) is
considered\footnote{Even more worrying, if we made use of $L_2- 2 L_1
  = {\cal O}(N_C^0)$ and replaced $(25 L_2^r(m_\rho) + 11 L_3)$ by
  $(50 L_1^r(m_\rho) + 11 L_3)$, within the same accuracy in the $N_C$
  expansion, this LEC's combination would come out now negative with
  the set of parameters of Eq.~(\ref{eq:Lr2's}), and substantially
  closer to zero than before, when the parameters of
  Eq.~(\ref{eq:Lr's}) are used.}.
\begin{itemize}
\item When $(25
L_2^r(m_\rho) + 11 L_3)$ is negative, the real part of the denominator
of Eq.~(\ref{eq:ssigma}) approaches  zero for increasing $N_C$,
which implies that $m_\sigma$ also approaches  zero, while the width
grows even faster that $m_\sigma^{-1}$.  From a given value of $N_C$
on, such that the real part of the denominator of
Eq.~(\ref{eq:ssigma}) becomes negative,  Eq.~(\ref{eq:ssigma}) does
not admit solution in the fourth quadrant. In this scenario, the
$\sigma-$resonance disappears, in the $N_C\gg 3$ limit, from this
quadrant of the SRS, and the pole appears in the third
quadrant\footnote{To compute the SRS amplitude at $s=-a-{\rm i}\ b$, 
with $a>0, b>0$ (third quadrant), 
Eqs.~(\ref{eq:srs1}) and~(\ref{eq:srs2}) should be replaced by 
\begin{eqnarray}
\ln(-s)&=&\ln|s|+{\rm i}\ {\rm arctan}\left(\frac{b}{a}\right)
 - 2\pi {\rm i}  \\
\ln(s)&=&\ln|s|+{\rm
  i}~{\rm arctan}\left(\frac{b}{a}\right)-{\rm i} \pi
\end{eqnarray}
\label{foot:a}},
though $\sqrt{s_\sigma}$ still lies in the fourth quadrant. This is
precisely the variable used in Refs.~\cite{Pelaez:2003dy,
Pelaez:2006nj}. To illustrate this point, in Fig.~\ref{fig:1} we show
results for the $\sigma-$pole, as a function of $N_C$. We have fixed
$L^r_{1,2,3}(m_\rho)$ to the values labeled as IAM in Table I of
Ref.~\cite{Pelaez:2003dy}. From $N_C=11$ on, the pole moves to the
third quadrant, while $\sqrt{s_\sigma}$ is still placed in the fourth
quadrant. These massless pion $\sqrt{s_\sigma}$ results  compare
nicely to those displayed in the middle panels of Fig.~2 of
Ref.~\cite{Pelaez:2003dy}, indicating that neglecting pion mass and
coupled channel effects constitutes also a good approximation to
analyze the large $N_C$ behaviour of this resonance. The behaviour at
small values of $N_C$ close to 3 support the conclusions of
Refs.~\cite{Pelaez:2003dy,Pelaez:2006nj} on the non-dominant $q\bar q$
nature of the $\sigma$ in the real world, while the results for larger
values of $N_C$ would indicate that there does not exist a  subdominant
$q\bar q$ component in the $\sigma$ wave function. Nevertheless, these
predictions are subject to sizable uncertainties beyond let us say
$N_C=10$, since $|\sqrt{s_\sigma}|$ is already 0.9 GeV for $N_C=10$,
and it reaches values close to 1.35 GeV for $N_C=30$.  The
applicability of the one-loop IAM is, at least, doubtful for these
large values of $\left |\sqrt{s}\right|$. Although the one loop IAM
amplitude incorporates contributions to all orders in the chiral
expansion (increasing powers of $1/f^{2n}$), it only accurately
accounts for those needed to exactly restore two-body elastic
unitarity. Therefore, a more precise knowledge of the leading $N_C$
terms of the ${\cal O}(p^6)$, ${\cal O}(p^8)$, etc... amplitudes seems
to be required. In this context the two loop IAM analysis carried out
in Ref.~\cite{Pelaez:2006nj} is better founded.
\begin{figure}[b]
\centerline{\hspace{-2cm}\includegraphics[height=8cm]{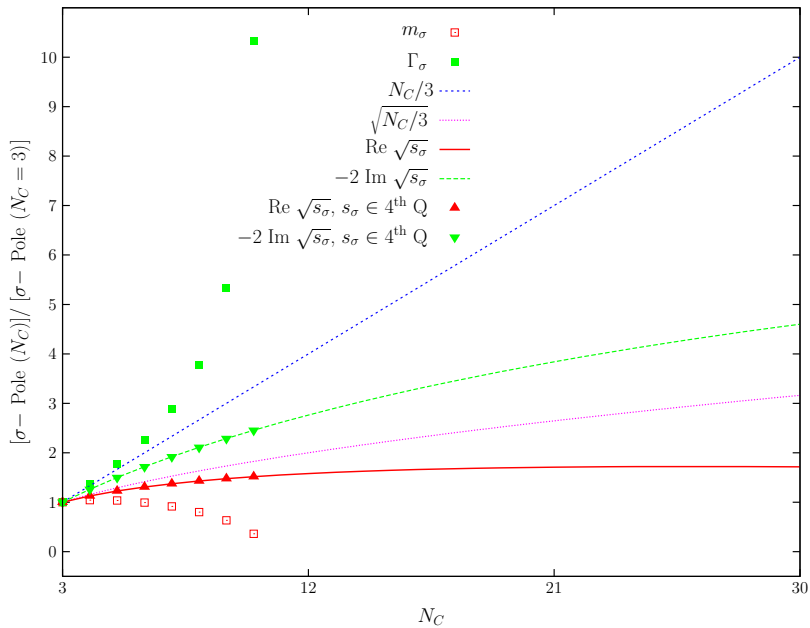}}
\caption{ $\sigma-$pole results, as a function of the number of
  colors, obtained with $L^r_{1,2,3}(m_\rho)$ fixed to the values
  labeled as IAM in Table I of Ref.~\cite{Pelaez:2003dy} and extended
  to arbitrary $N_C$ as in that reference. Up to $N_C=10$ we
  solve Eq.~(\ref{eq:ssigma}), beyond this number of colors, the pole
  lies in the third quadrant (we solve Eq.~(\ref{eq:ssigma}) with the
  modifications mentioned in footnote \ref{foot:a}). All results are
  normalized by the $N_C=3$ results: $m_\sigma= 328.1$ MeV,
  $\Gamma_\sigma=629.6$ MeV and $\sqrt{s_\sigma}= \left( 412.6 - {\rm
  i}\ 250.3 \right)$ MeV.}
\label{fig:1}
\end{figure}
\item On the contrary, when the combination $(25 L_2^r(m_\rho) + 11
L_3)$ is positive, the large $N_C$ limit of the $\sigma-$meson is
qualitatively identical to that of the $\rho-$meson, with $\widehat
m_\sigma \sim {\cal O}(N_C^0)$ and $\widehat \Gamma_\sigma \sim {\cal
O}(N_C^{-1})$. There will exist a
$q\bar q$ component in the $\sigma$ meson that  would become dominant
in the $N_C\gg 3$ limit. 

However, if it happens that for $N_C=3$, the magnitude $192\pi^2(25
L_2^r(m_\rho) + 11 L_3)$, though positive, is close to zero and
significantly smaller than 51/2, there exists a transient region of
low and intermediate values of $N_C$ where the expected scaling rules
are not satisfied and the chiral log $-25\ln\left(
s_\sigma/\mu^2\right)$ induces non trivial and unexpected dependences
on $N_C$.  This indicates that the $q\bar q$ component of the $\sigma$
is sub-dominant when $N_C=3$. Once $N_C$ is sufficiently large, the
real part becomes rather constant, while the imaginary part starts
decreasing, as predicted by Eq.~(\ref{eq:pole-sigma}). However, this
asymptotic value for the mass would be out of the range of
applicability of one loop ChPT, and the caveats mentioned above on the
need of some detailed leading $N_C$ ${\cal O}(p^6)$, ${\cal
O}(p^8)$,.. input will apply also here.
\end{itemize}
Thus, the $N_C\gg 3$ behaviour of the $\sigma-$meson within the IAM
method to one loop depends critically on the sign and size of a
parameter combination which value cannot be pinned down reliably with
the needed accuracy. This crucial role played by the critical LEC's
combination $(25 L_2^r(m_\rho) + 11 L_3)$ to determine the $N_C\to
\infty$ limit of the $\sigma-$resonance was firstly pointed out
in~\cite{Sun:2005uk}.  Here, we provide in addition an error analysis,
address the relevance of identifying  the leading $N_C$ term
in the decomposition of Eq.~(\ref{eq:eles-nc}), analyze the
relation between $\sigma$ and $\rho$ channels (see below also) and
discuss the transition to a region where the pole is located in  ${\rm Re}\ s
<0$ half-plane\footnote{Note that in this case,
the path integral for the resonance field would not be  well
  defined.}. The above discussion clearly
illustrates the need of having some reliable and independent
insight\footnote{Though desirable, the knowledge of such a
decomposition for $f$ is less relevant, since it will not affect 
the existence or not of the $\sigma$ state in the $N_C\gg 3$ limit.}  into
the $N_C$ leading and sub-leading terms of Eq.~(\ref{eq:eles-nc}). To
this end, we will next make use of the resonance saturation
approach~\cite{Ecker:1988te,Ecker:1989yg}.

\section{LEC's: Resonance Saturation Approach 
and Large $N_C$ Limit}

In the resonance saturation approach one writes down a Lagrangian
including the resonance fields and integrate them
out~\cite{Ecker:1988te,Ecker:1989yg}, yielding values for the LEC's at
some scale not too far away from the resonance region. It is common
practice to adopt $\mu=m_\rho$ as a reasonable choice. The
generalization of this approach to the large $N_C$ limit requires, in
addition to including infinitely many resonances, the use of short
distance constraints~\cite{Pich:2002xy}, which are conditions stemming
from the analysis of Green's functions in QCD at high momentum.  In
the Single Resonance Approximation (SRA), each infinite resonance sum
is approximated with the contribution from the first meson nonet with
the given quantum numbers. This is meaningful at low energies where
the contributions from higher--mass states are suppressed by their
corresponding propagators. Within SRA one finds (see
Refs.~\cite{Ecker:1988te,Ecker:1989yg} for notation),
\begin{eqnarray}
\left[ L_1^r(m_\rho)\right]^{\rm SRA} &=& \frac{G_V^2}{8 M_V^2} -
  \frac{c_d^2}{6M_S^2}+ \frac{\tilde c_d^2}{2M_{S_1}^2}  \label{eq:l1-sra}\\ 
\left [ L_2^r(m_\rho)\right]^{\rm SRA} &=& \frac{G_V^2}{4 M_V^2}
  \label{eq:l2-sra} \\ 
L_3^{\rm SRA} &=& - \frac{3G_V^2}{4M_V^2} + \frac{c_d^2}{2M_S^2} 
\label{eq:l3-sra}
\end{eqnarray}
where $M_{S_1}$ and $M_S$ are the singlet and octet scalar masses,
respectively and $M_V$ that of the nonet of vector mesons. 

In the  $N_C\gg 3$ limit, octet and singlet mesons become degenerate
and thus $M_{S_1}=M_S$, while $G_V$, $c_d$ and $\tilde c_d$ are all
${\cal O}(N_C^\frac12)$. The large $N_C$ condition
$L_2- 2 L_1 = {\cal O} (N_C^0)$ can then be achieved by taking $c_d^2=
3 \tilde c^2_d$, while the short distance constraints can be used to
determine the resonance couplings at leading order in the $N_C$
expansion~\cite{Ecker:1988te,Ecker:1989yg,Pich:2002xy},
\begin{equation}
\sqrt{2} G_V = 2 c_d = f \label{eq:short-distances}
\end{equation}
All together this allows to estimate the leading $N_C$ terms
($A_i's$ coefficients in Eq.~(\ref{eq:eles-nc})) of $L_2^r(m_\rho)$
and $L_3$ LEC's~\cite{Pich:2002xy},
\begin{equation}
[L_2^r(m_\rho)]^{\rm SRA} = \frac{f^2}{8M_V^2}+{\cal O}(N_C^0), 
\qquad L_3^{\rm SRA} = -\frac{3f^2}{8M^2_V} +
\frac{f^2}{8M^2_S} +{\cal O}(N_C^0) \label{eq:eles-sra-analitica}
\end{equation}
The subleading $N_C$ corrections to the SRA predictions of the LEC's are
difficult to estimate in a model independent way (see however
\cite{Rosell:2004mn, Xiao:2007pu}). The last column of Table 2 in
Ref.~\cite{Pich:2002xy} shows\footnote{Note that in this reference,
$f$ is fixed to 92 MeV.} the estimates for $L_2^r(m_\rho)$ and $L_3$
obtained with $M_V=0.77$ GeV and $M_S=1.0$ GeV, and those, by simply
scaling with $N_C/3$, can be used to determine $\widehat L_{2,3}$,
\begin{equation}
10^3\ \widehat L_2^{\rm SRA} = 1.8 \times \frac{N_C}{3}, 
\qquad 10^3\ \widehat L_3^{\rm
  SRA} = -4.3 \times \frac{N_C}{3}  \label{eq:eles-sra}
\end{equation}
The above value for $L_3^{\rm SRA}$ provides an estimate for the
$\rho-$mass, in the $N_C\gg 3$ limit, of around 700 MeV, while its width
decreases as $1/N_C$ [Eq.~(\ref{eq:pole-rho})]. This result is in good
agreement with the findings of Ref.~\cite{Pelaez:2003dy}, also
obtained within the one loop IAM scheme, but including both finite
pion mass and coupled channel effects. The results for the
$\rho-$resonance are robust, and since the mass is moderately small,
we expect they would not be much affected by leading $N_C$
contributions showing up at order ${\cal O}(p^6)$ or higher, as 
confirmed by the two loop results of
Ref.~\cite{Pelaez:2006nj}. Under these circumstances, if we identify
$\widehat m_\rho$ with $M_V$ (mass of the nonet of vector mesons that
is introduced in the SRA), and require consistency between
Eqs.~(\ref{eq:pole-rho}) and~(\ref{eq:eles-sra-analitica}), we 
find
\begin{equation}
 L_3^{\rm SRA} = -\frac{3f^2}{8M^2_V} +
\frac{f^2}{8M^2_S}  +{\cal O}(N_C^0) =  -\frac{ f^2}{4 M_V^2}\left (1 +{\cal
  O}(m^2/M_V^2)\right) + {\cal O}(N_C^0)
\end{equation}
where we have used that $\widehat f^2/\widehat L_3 = f^2/L_3+{\cal
  O}(1/N_C)$. From the above equation, it trivially follows 
\begin{equation}
M_S=M_V +{\cal O}(1/N_C) + {\cal O}(mN_C^0 )\label{eq:mvs}
\end{equation}
which is based on the relation $G_V=\sqrt2 c_d+{\cal O}(N_C^0)$
[Eq.~(\ref{eq:eles-sra-analitica})] and the assumed scheme, namely 
SRA--one--loop IAM. It is worth noting that the large $N_c$ identity of
scalar and vector meson masses, Eq.~(\ref{eq:mvs}), has also been
derived within the context of mended symmetries~\cite{Weinberg:1990xn}
as well as chiral quark models~\cite{Megias:2004uj}. Computing finite
pion mass corrections to Eq.~(\ref{eq:mvs}) is
straightforward. Indeed, in the large $N_C$ limit, the only relevant
pion mass corrections to the amplitude are those proportional to
${\bar l}_1-{\bar l}_2$ and $\bar l_4$ (see for instance Eq.~(B7) in
Ref.~\cite{Nieves:1999bx}). Taking into account that $\bar l_4$ is
determined by the combination $2L_4^r(\mu)+L_5^r(\mu)$ and that $L_5$
dominates in the $N_C\gg 3$ limit, one easily finds
\begin{equation}
  \widehat m_\rho^2 = -\frac{\widehat f^2}{4\widehat
      L_3}\left(1-\frac{8m^2}{\widehat f^2}\widehat L_5
      \right) 
\end{equation}
and by using~\cite{Pich:2002xy},
\begin{equation}
[L_5^r(m_\rho)]^{\rm SRA} = \frac{f^2}{4M_S^2}+{\cal O}(N_C^0),
\end{equation} 
the SRA--one--loop IAM consistency requirement now leads to
\begin{equation}
M_S^2=M_V^2-4m^2 +{\cal O}(1/N_C) \label{eq:mvs2}
\end{equation}
which constitutes one of the main results of this work.

The situation, however, is totally different in the scalar
sector. From the estimates given in Eq.~(\ref{eq:eles-sra}), we find
that the combination $(25 \widehat L_2^{\rm SRA} + 11 \widehat
L_3^{\rm SRA})$ turns out to be small compared to the log contribution
and negative.  Thus, for $N_C$ sufficient large the $\sigma-$pole
disappears from the fourth quadrant of the SRS, and one finds a
similar behaviour to that depicted in Fig.~\ref{fig:1}.  This might
hint at a definite non $q\bar q$ nature of the $\sigma-$resonance, as
suggested in Ref.~\cite{Pelaez:2003dy}, since there is no trace even
of the existence of a sub-dominant $q\bar q$ component. Finite pion
mass corrections can be easily taken into account, but they do not
modify the discussion.  Note however, once more, the large
cancellation that occurs in this combination of LEC's, since $|(25
\widehat L_2^{\rm SRA} + 11 \widehat L_3^{\rm SRA})/(11\widehat
L_3^{\rm SRA})| = 0.05$. As a consequence, it is difficult to draw any
robust conclusion on the nature (existence or not of the $q\bar q$
component) of the lightest spin--isospin scalar meson in the $N_C \to
\infty $ limit.  Actually, a small variation of the short distance
constraint relations of Eq.~(\ref{eq:short-distances}), increasing
slightly the ratio $c_d/G_V$, or approximating $M_S$ by $M_V\sim 770$
MeV, as suggested by Eq.~(\ref{eq:mvs}), would reduce $|L_3^{\rm
SRA}|$ and lead to positive values for the LEC's combination $(25
\widehat L_2^{\rm SRA} + 11 \widehat L_3^{\rm SRA})$.

Under these circumstances, we will consider a different scenario. Let
us assume that $(25 \widehat L_2^{\rm SRA} + 11 \widehat L_3^{\rm
  SRA})$ is positive, leading then to a stable $\sigma-$resonance in
the $N_C\gg 3$ limit. Its mass then could be identified to that of the
first nonet of scalar mesons introduced in the SRA. If we take 
$M_S=\widehat m_\sigma$ and $M_V=\widehat m_\rho$, and requiring
consistency, between Eqs.~(\ref{eq:l2-sra}--\ref{eq:l3-sra}) and the
mass of the $\sigma-$ and $\rho-$mesons in the large $N_C$ limit, given
in Eqs.~(\ref{eq:pole-rho}--\ref{eq:pole-sigma}), we find
\begin{eqnarray}
M_S = 2 M_V +{\cal O}(1/N_C), \qquad c_d = f
+{\cal O}(1/N_C) 
\end{eqnarray}
for massless pions. The second condition ($c_d \sim f$) is
phenomenologically strongly disfavored~\cite{Pich:2002xy}, while the
first one contradicts the more robust result of Eq.~(\ref{eq:mvs}) (or
~(\ref{eq:mvs2})) based only in the vector channel.  There are
different ways to circumvent this apparent contradiction: not identifying
$M_S=\widehat m_\sigma$ and assuming that, in the large $N_C$ limit, the
$\sigma$ meson becomes heavier than the scalar meson nonet, invoked in
the SRA, or correcting  the SRA estimates of the LEC's by considering, 
for instance,  contributions of tensor
resonances~\cite{Donoghue:1988ed}, etc....  Nevertheless, one should
bear in mind that for this large
value of the mass ($\sim 2 M_V$), IAM results, based on the first two
orders of the chiral expansion, should not be  very reliable because of
the limited control on the leading $N_C$ terms appearing beyond
$1/f^4$.

\section{Conclusions}

We have started looking at the chiral limit of the one loop SU(2) ChPT
amplitudes. We have shown how in the chiral limit, a major source of
distinction between the $\sigma-$ and $\rho-$ channels is due to the
role played by chiral logarithms; while for the $\sigma-$channel the
logs add up into sizeable contributions,  in the $\rho-$channel they
cancel exactly.  Next, we have used the IAM to unitarize the
amplitudes and have looked for poles in the SRS of
the amplitudes. We have found a fair description of the established
properties of the $\sigma-$ and $\rho-$resonances, showing the little
effect of finite pion mass and coupled channel corrections.

The properties of these resonances for growing number of colors have
been also discussed. Our results confirm and explain the different
behavior, in agreement with Refs.~\cite{Pelaez:2003dy, Pelaez:2006nj}, of the
$\rho$ and $\sigma$ for $N_C$ not far from 3, but, when extending the
analysis to the large $N_C$ limit, no robust conclusion of the
$\sigma$ pole behavior can be inferred from the one loop IAM
only. This is due to the large cancellation existing in the
combination of LEC's that govern the scalar channel; there exists a
critical value for a combination of LEC's which cannot be pinned down
with the needed accuracy at {\it any} value of $N_C$.

Finally, we have discussed further constraints deduced by requiring
consistency between resonance saturation and unitarization. By looking
only at the $\rho-$channel, we have found that the masses of the first
vector and scalar meson nonets, invoked in the SRA, turn out to be
degenerated in the large $N_C$ limit (see Eq.~(\ref{eq:mvs}) or
Eq.~(\ref{eq:mvs2}) that incorporates finite pion mass
corrections). The two loop calculation of Ref.~\cite{Pelaez:2006nj} supports
 this latter result. If we look at the right top panel of
Fig. 1 of this reference, we observe that above the $N_C=6-10$ region,
the $\sigma-$mass becomes rather constant, while its width rapidly
decreases. Moreover, this large $N_C$ asymptotic value of the mass is
around two or three times bigger than ${\rm Re}\ \sqrt{s_\sigma}\Big
|_{N_C=3}$, and thus it lies in the 1 GeV region, quite close then to
the $\rho-$mass. 

Within the restricted IAM unitarization approach assumed here, we
would find a scenario where the $\sigma$ resonance would become
degenerate with the $\rho$ and stable, for a sufficiently large number
of colors. This complies to the consequences of mended
symmetries~\cite{Weinberg:1990xn} as well as with chiral quark model
calculations~\cite{Megias:2004uj}. However, the nature of the
$\sigma-$resonance in the real world ($N_C=3$) would be totally
different to that of the $\rho-$meson, being it mostly governed by
chiral logarithms stemming from unitarity and crossing symmetry,
justifying the widely accepted nature of the $\sigma$ as a dynamically
generated meson.

\begin{acknowledgments} 
   We warmly thank J.R. Pel\'aez, E. Oset, A. Pich and
   M.J. Vicente--Vacas for useful discussions. This research was
   supported by DGI and FEDER funds, under contracts FIS2008-01143/FIS
   and the Spanish Consolider-Ingenio 2010 Programme CPAN
   (CSD2007-00042), by Junta de Andaluc\'\i a under contract FQM0225,
   and it is part of the European Community-Research Infrastructure
   Integrating Activity ``Study of Strongly Interacting Matter''
   (acronym HadronPhysics2, Grant Agreement n. 227431) and of the EU
   Human Resources and Mobility Activity ``FLAVIAnet'' (contract
   number MRTN--CT--2006--035482) , under the Seventh Framework
   Programme of EU.

\end{acknowledgments}


\end{document}